\documentclass[english,american]{article}
\usepackage[T1]{fontenc}
\usepackage[utf8]{inputenc}
\usepackage[skip=\baselineskip]{parskip}
\usepackage{booktabs}
\usepackage{amsmath}
\usepackage{amssymb}
\usepackage{setspace}
\makeatletter

\providecommand{\tabularnewline}{\\}

\makeatother

\usepackage{babel}
\begin{document}
\title{A Corrected Welch--Satterthwaite Equation. And: What You Always Wanted
to Know About Kish's Effective Sample Size but Were Afraid to Ask.}
\author{Matthias von Davier}
\maketitle
\begin{abstract}
This article presents a corrected version of the Satterthwaite (1941,
1946) approximation for the degrees of freedom of a weighted sum of
independent variance components. The original formula is known to
yield biased estimates when component degrees of freedom are small.
The correction, derived from exact moment matching, adjusts for the
bias by incorporating a factor that accounts for the estimation of
fourth moments. We show that Kish's (1965) effective sample size formula
emerges as a special case when all variance components are equal and
component degrees of freedom are ignored. Simulation studies demonstrate
that the corrected estimator closely matches the expected degrees
of freedom even for small component sizes, while the original Satterthwaite
estimator exhibits substantial downward bias. Additional applications
are discussed, including jackknife variance estimation, multiple imputation
total variance, and the Welch test for unequal variances.
\end{abstract}

\section{Introduction}

The estimation of effective degrees of freedom for complex variance
estimators is a recurring problem in statistics, particularly when
combining independent variance components with different degrees of
freedom. Satterthwaite (1941, 1946) provided a widely used approximation
that expresses the degrees of freedom of a weighted sum of variance
components as a function of the component variances and their degrees
of freedom. This approximation underpins many statistical procedures,
including the Welch test for two-sample comparisons with unequal variances,
and variance estimation in complex survey designs. However, the Satterthwaite
formula relies on large-sample approximations and can be severely
biased when the component degrees of freedom are small or the number
of components is limited (Johnson \& Rust, 1992).

In parallel, Kish (1965) introduced the concept of effective sample
size for weighted samples, which adjusts the nominal sample size to
account for unequal weighting. Kish's formula, derived under the assumption
of equal variances across observations, has become a standard tool
in survey sampling. Although developed in different contexts, both
Satterthwaite's and Kish's estimators share a common structure and
can be linked through a simplified Satterthwaite expression when component
variances are homogeneous.

This article has three main objectives. First, we derive a corrected
Satterthwaite degrees\nobreakdash-of\nobreakdash-freedom estimator
that mitigates the bias for small component degrees of freedom. The
correction follows from a careful handling of the expected fourth
moments of the component variance estimates, leading to an improved
formula that reduces to the original Satterthwaite expression as component
degrees of freedom increase. Second, we show that Kish's effective
sample size is a special case of the Satterthwaite estimator when
all variance components are equal and the component degrees of freedom
are effectively ignored. Third, we present simulation results comparing
the original Satterthwaite estimator, the corrected version, and Kish's
estimator under various conditions, demonstrating the bias reduction
achieved by the correction. Finally, we illustrate applications of
the corrected formula in jackknife variance estimation, multiple imputation
total variance, and the Welch test.

The article is organized as follows. Section 2 reviews Satterthwaite's
original derivation and presents the corrected formula. Section 3
discusses Kish's effective sample size and its relationship to Satterthwaite's
estimator. Section 4 provides simulation evidence comparing the three
estimators. Section 5 gives examples of the corrected estimator beyond
effective sample size, including jackknifing, multiple imputation,
and the Welch test. Section 6 concludes with a summary of recommendations.

\section{Satterthwaite's (1941, 1946) Approximate d.f.}

Satterthwaite (1941, 1946) examines the approximate degrees of freedom
for complex variance estimates. That is, we have weights $w_{1},\dots,w_{k}$and
variance components $S_{1}^{2},\dots S_{k}^{2}$, each with degrees
of freedom $\nu_{k}$, so that 
\[
\frac{\nu_{k}S_{k}^{2}}{\sigma_{k}^{2}}\sim\chi_{\nu_{k}}^{2}
\]
that is,
\[
E\left(\frac{\nu_{k}S_{k}^{2}}{\sigma_{k}^{2}}\right)=\nu_{k}\:\mathrm{and}\:Var\left(\frac{\nu_{k}S_{k}^{2}}{\sigma_{k}^{2}}\right)=2\nu_{k}.
\]

The approximate d.f. are developed for a complex variance estimate
\[
S^{2}=\sum_{k=1}^{K}w_{k}S_{k}^{2}
\]
which is a linear combination of independent variance components.
Satterthwaite (1941, 1946) argues that with independent $X_{k}$ and
reasonably large d.f. for each of the variance components $S_{k}^{2}$,
one may assume that
\[
S^{2}\sim\chi_{\nu^{*}}^{2}
\]
and that 

\[
d.f.\left(S^{2}\right)=\nu^{*}\approx\frac{\left(\sum_{i}w_{i}S_{i}^{2}\right)^{2}}{\sum_{i}w_{i}^{2}\frac{S_{i}^{4}}{\nu_{k}}}.
\]

\subsection{Proof of Satterthwaite's Formula and its Improvement}

The formula provided by Satterthwaite (1941, 1946) is widely used,
but it is known that it produces biased estimated (e.g. Johnson \&
Rust, 1992) for small component degrees of freedom $\nu_{k}$. In
this section, we derive the expected value the approximation is based
on and provide a tighter approximation suitable for small $\nu_{k}$
and $K$, while it converges to the original for large $K$ and large
$\nu_{k}$.

This section starts with reviewing some of the properties of $\chi^{2}$distributed
variables that are needed to derive the main result.

\subsubsection{Some Properties of $\chi^{2}$ Distributed Random Variables}

Note that for any $\chi^{2}$ distributed variable with $\nu$ degrees
of freedom we have
\[
\frac{\nu^{2}Var\left(S^{2}\right)}{\sigma^{4}}=2\nu\rightarrow\nu=2\frac{\left[\sigma^{2}\right]^{2}}{Var\left(S^{2}\right)}
\]

and
\[
\frac{\nu^{2}Var\left(S^{2}\right)}{\sigma^{4}}=2\nu\rightarrow Var\left(S^{2}\right)=2\frac{\left[\sigma^{2}\right]^{2}}{\nu}.
\]

Also, since $Var\left(X\right)=E\left(X^{2}\right)-E\left(X\right)^{2}$we
have $Var\left(X\right)+E\left(X\right)^{2}=E\left(X^{2}\right)$.
Then we can continue with plugging in $S^{2}$ for $X$ and find 
\[
Var\left(S^{2}\right)+E\left(S^{2}\right)^{2}=E\left(S^{4}\right)=2\frac{\sigma^{4}}{\nu}+\sigma^{4}=\sigma^{4}\left(\frac{2}{\nu}+1\right)=\sigma^{4}\left(\frac{2+\nu}{\nu}\right)
\]
and hence 
\[
\sigma^{4}=E\left(S^{4}\right)\frac{\nu_{S}}{\nu_{S}+2}.
\]

\subsubsection{Derivation of the Satterthwaite Formula}

Since the $S_{k}^{2}$ are independent we have $Var\left(S^{2}\right)=\sum_{k}w_{k}^{2}Var\left(S_{k}^{2}\right)$
and by using the equivalence $\nu=2\frac{\left[\sigma^{2}\right]^{2}}{Var\left(S^{2}\right)}\leftrightarrow Var\left(S^{2}\right)=2\frac{\left[\sigma^{2}\right]^{2}}{\nu}$
for $S^{2}$and $S_{k}^{2}$ we obtained
\[
2\frac{\sigma^{4}}{\nu^{*}}=\sum_{k=1}^{K}w_{k}^{2}2\frac{\sigma_{k}^{4}}{\nu_{k}}
\]
which, by solving for $\nu^{*},$provides
\[
\nu^{*}=\frac{\sigma^{4}}{\sum_{k=1}^{K}w_{k}^{2}\frac{\sigma_{k}^{4}}{\nu_{k}}}
\]

Then, an estimate of $\nu^{*}$ is obtained by using the equivalence
and the approximation for large $\nu$
\[
\sigma^{4}=E\left(S^{4}\right)\frac{\nu}{\nu+2}\approx S^{4}
\]
both for $\sigma_{k}^{4}=\left(\sigma_{k}^{2}\right)^{2}=\left[E\left(S_{k}^{2}\right)\right]^{2}$
and $\sigma^{4}.$ This assumes that the components degrees of freedom
are sufficiently large so that $\frac{\nu}{\nu+2}\approx1$ and that
$S^{4}\approx E\left(S^{4}\right)$ which also is implied by $\nu\rightarrow\infty.$

Then, plugging in $S^{4}$ for $\sigma^{4}$, as well as $S_{k}^{2}$
for $\sigma_{k}^{2}$ yields 
\[
\nu^{*}\approx\frac{\left[\sum_{k}w_{k}S_{k}^{2}\right]^{2}}{\sum_{k}w_{k}^{2}\frac{\left[S_{k}^{2}\right]^{2}}{\nu_{k}}}.
\]

Satterthwaite (1941) points out this approximation may not work well
for small $K,\nu_{k}$.

\subsubsection{Some Equivalencies}

Here, it is important to note that 
\[
v_{satt}^{*}=\frac{\left[\sum_{k}w_{k}\sigma_{k}^{2}\right]^{2}}{\sum_{k}w_{k}^{2}\frac{\left[\sigma_{k}^{2}\right]^{2}}{\nu_{k}}}=\frac{\left[\sum_{k}cw_{k}\sigma_{k}^{2}\right]^{2}}{\sum_{k}\frac{\left[cw_{k}\sigma_{k}^{2}\right]^{2}}{\nu_{k}}}
\]
for any $c>0.$ Choosing 
\[
c=K\left[\sum_{k}w_{k}\sigma_{k}^{2}\right]^{-1}
\]
and setting $\overline{w\sigma^{2}}=\frac{1}{K}\sum_{k}w_{k}\sigma_{k}^{2}$
yields
\[
\nu_{satt}^{*}=\frac{\left[\sum\frac{w_{k}\sigma_{k}^{2}}{\overline{w\sigma^{2}}}\right]^{2}}{\sum_{k}\frac{1}{\nu_{k}}\left[\frac{w_{k}\sigma_{k}^{2}}{\overline{w\sigma^{2}}}\right]^{2}}=\frac{K^{2}}{\sum_{k}\frac{1}{\nu_{k}}\left[\frac{w_{k}\sigma_{k}^{2}}{\overline{w\sigma^{2}}}\right]^{2}}
\]
which is $K$ times the harmonic mean of the $q_{k}=\nu_{k}\left[\frac{\overline{w\sigma^{2}}}{w_{k}\sigma_{k}^{2}}\right]^{2}$
\[
\nu_{satt}^{*}=K\cdot H\left(q_{k}\right)=K\cdot H\left(\nu_{k}\left[\frac{\overline{w\sigma^{2}}}{w_{k}\sigma_{k}^{2}}\right]^{2}\right).
\]

This section shows that the neither the original Satterthwaite estimator
nor its corrected version depend on the average value $\overline{w\sigma^{2}}$
of the weighted variances $w_{i}\sigma_{k}^{2}$.

\subsubsection{Improvement for Small $\nu_{k}$ and Small $K$}

While Satterthwaite assumes that $K,\nu_{k}>>1$ and hence, for all
instances, his work implicitly uses $\frac{\nu}{\nu+2}\approx1$,
and then replaces $\sigma_{k}^{2}$ by $S_{k}^{2}$ directly, without
the factor for estimation of $\nu_{satt}^{*}$, von Davier (2025a,b)
points out that for small $K,\nu_{k}$ a correction is needed. Recall
that we have shown above that 
\[
\hat{\sigma}_{k}^{4}\approx S_{k}^{4}\frac{\nu_{k}}{\nu_{k}+2}\;\mathrm{and}\;\sigma^{4}\approx S^{4}\frac{\nu^{*}}{\nu^{*}+2}
\]
which provide the necessary correction. It follows that 

\[
\nu^{*}\approx\frac{\frac{\nu^{*}}{\nu^{*}+2}\left[\sum_{k}w_{k}S_{k}^{2}\right]^{2}}{\sum_{k}w_{k}^{2}\frac{\left[S_{k}^{2}\right]^{2}}{\nu_{k}+2}}\leftrightarrow\nu^{*}\approx\frac{\left[\sum_{k}w_{k}S_{k}^{2}\right]^{2}}{\sum_{k}w_{k}^{2}\frac{\left[S_{k}^{2}\right]^{2}}{\nu_{k}+2}}-2.
\]

Interestingly, Boardman (1974) discusses a similar estimator for $K=2$,
\[
\nu^{*}\approx\frac{\left(a_{1}S_{1}^{2}+a_{2}S_{2}^{2}\right)^{2}}{\frac{a_{1}^{2}S_{1}^{4}}{\nu_{1}+2}=\frac{a_{2}^{2}S_{2}^{4}}{\nu_{2}+2}}
\]
which he refers to as the 'unbiased' Satterthwaite, but does not appear
to derive the result, an neither includes a reference. However, he
mentions that Satterthwaite (1941) was aware that his formula is biased
for small $\nu_{k}$. Note that Boardman's (1974) estimator is missing
the final term $-2$ that corrects the estimator. 

\subsection{Some Results on the Improved Effective D.F. }

The following tables show results for sums of $K$ simple variance
estimates $S_{k}^{2}$ with $\frac{\nu_{k}S_{k}^{2}}{\sigma_{k}^{2}}=X_{k,\nu_{k}}^{2}\sim\chi_{\nu_{k}}^{2}.$
The ideal case where $\forall k,j:\sigma_{k}^{2}=\sigma_{j}^{2}=\sigma^{2},\nu_{k}=\nu_{j}=\overline{\nu},w_{k}=w_{j}=\frac{1}{K}$
so that $Kw_{i}=1$ was used to enable comparisons of the original
Satterthwaite equation to the improved estimator, and against the
expected value 
\[
E\left(\nu^{*}\right)=K\overline{\nu}
\]
that follows in this case since $\frac{K\overline{\nu}}{\sigma^{2}}S^{2}=\sum_{k=1}^{K}Kw_{i}\frac{\overline{\nu}}{\sigma^{2}}S_{k}^{2}=\sum_{k=1}^{K}X_{k,\overline{\nu}}^{2}\sim\chi_{K\overline{\nu}}^{2}.$ 

\selectlanguage{english}%
\begin{table}[htbp]
\centering \caption{Simulation results for $K=2$ and $K=4$}
\begin{tabular}{rrrrrrr}
\toprule 
$K$ & $df$ & $\bar{\nu}_{\text{unc}}$ & SD unc & $\bar{\nu}_{\text{corr}}$ & SD corr & $K\times\bar{\nu}$\tabularnewline
\midrule 
2 & 1 & 1.410 & 0.346 & 2.229 & 1.039 & 2\tabularnewline
2 & 2 & 3.135 & 0.640 & 4.270 & 1.281 & 4\tabularnewline
2 & 4 & 6.837 & 1.044 & 8.256 & 1.566 & 8\tabularnewline
2 & 8 & 14.577 & 1.494 & 16.222 & 1.868 & 16\tabularnewline
2 & 16 & 30.348 & 1.938 & 32.141 & 2.181 & 32\tabularnewline
2 & 32 & 62.199 & 2.282 & 64.086 & 2.425 & 64\tabularnewline
\midrule 
4 & 1 & 2.198 & 0.637 & 4.595 & 1.912 & 4\tabularnewline
4 & 2 & 5.320 & 1.202 & 8.640 & 2.403 & 8\tabularnewline
4 & 4 & 12.400 & 1.974 & 16.600 & 2.961 & 16\tabularnewline
4 & 8 & 27.567 & 2.852 & 32.459 & 3.565 & 32\tabularnewline
4 & 16 & 58.908 & 3.567 & 64.272 & 4.013 & 64\tabularnewline
4 & 32 & 122.482 & 4.113 & 128.138 & 4.370 & 128\tabularnewline
\bottomrule
\end{tabular}
\end{table}

\begin{table}[htbp]
\centering \caption{Simulation results for $K=8$ and $K=16$}
\begin{tabular}{rrrrrrr}
\toprule 
$K$ & $df$ & $\bar{\nu}_{\text{satt}}$ & SD unc & $\bar{\nu}_{\text{corr}}$ & SD corr & $K\times\bar{\nu}$\tabularnewline
\midrule 
8 & 1 & 3.656 & 1.023 & 8.969 & 3.070 & 8\tabularnewline
8 & 2 & 9.473 & 2.014 & 16.946 & 4.029 & 16\tabularnewline
8 & 4 & 23.171 & 3.296 & 32.757 & 4.944 & 32\tabularnewline
8 & 8 & 53.198 & 4.720 & 64.498 & 5.900 & 64\tabularnewline
8 & 16 & 115.997 & 5.736 & 128.497 & 6.453 & 128\tabularnewline
8 & 32 & 243.096 & 6.552 & 256.290 & 6.962 & 256\tabularnewline
\midrule 
16 & 1 & 6.443 & 1.608 & 17.330 & 4.825 & 16\tabularnewline
16 & 2 & 17.586 & 3.180 & 33.172 & 6.360 & 32\tabularnewline
16 & 4 & 44.642 & 5.226 & 64.963 & 7.838 & 64\tabularnewline
16 & 8 & 104.373 & 7.278 & 128.466 & 9.097 & 128\tabularnewline
16 & 16 & 229.741 & 8.776 & 256.458 & 9.873 & 256\tabularnewline
16 & 32 & 484.036 & 9.730 & 512.289 & 10.338 & 512\tabularnewline
\bottomrule
\end{tabular}
\end{table}

\begin{table}[htbp]
\centering \caption{Simulation results for $K=32$ and $K=64$}
\begin{tabular}{rrrrrrr}
\toprule 
$K$ & $df$ & $\bar{\nu}_{\text{satt}}$ & SD unc & $\bar{\nu}_{\text{corr}}$ & SD corr & $K\times\bar{n}$\tabularnewline
\midrule 
32 & 1 & 11.938 & 2.503 & 33.815 & 7.509 & 32\tabularnewline
32 & 2 & 33.825 & 4.924 & 65.650 & 9.847 & 64\tabularnewline
32 & 4 & 87.394 & 7.914 & 129.091 & 11.870 & 128\tabularnewline
32 & 8 & 207.089 & 10.860 & 256.861 & 13.575 & 256\tabularnewline
32 & 16 & 457.054 & 13.031 & 512.186 & 14.659 & 512\tabularnewline
32 & 32 & 965.813 & 14.191 & 1024.177 & 15.078 & 1024\tabularnewline
\midrule 
64 & 1 & 22.725 & 3.757 & 66.175 & 11.270 & 64\tabularnewline
64 & 2 & 65.904 & 7.259 & 129.807 & 14.517 & 128\tabularnewline
64 & 4 & 172.991 & 11.588 & 257.486 & 17.381 & 256\tabularnewline
64 & 8 & 412.037 & 15.757 & 513.047 & 19.696 & 512\tabularnewline
64 & 16 & 912.294 & 18.557 & 1024.331 & 20.876 & 1024\tabularnewline
64 & 32 & 1929.855 & 20.471 & 2048.471 & 21.750 & 2048\tabularnewline
\bottomrule
\end{tabular}
\end{table}

\selectlanguage{american}%
The obvious conclusion is that the correction of the Satterthwaite
equation for estimating effective sample size is necessary even for
reasonably large component degrees of freedom. The case where $K=64$
and $\nu_{k}=\overline{\nu}=32$ shows this clearly. The average $df$
for the corrected estimate is $\overline{\nu_{corr}^{*}}=2048.5\approx2028=K\times\overline{\nu}>\overline{\nu_{satt}^{*}}=1929.9$.
The other end of the table shows for $K=2$ and $df=1$ that $\overline{\nu_{corr}^{*}}=2.23\approx2=K\times\overline{\nu}>1.41=\overline{\nu_{satt}^{*}}.$

\section{Kish's (1965) Effective Sample Size Estimate}

To start with the similarities, both Kish's (1965) formula for the
effective sample size for an estimator of the variance of the mean
and the Satterthwaite (1941) estimator of effective degrees of freedom
of a complex variance estimate conceptual similarities. 

The idea behind finding the effective sample size for a weighted sample
mean can be illustrated using the case where some groups do not carry
any weight, i.e., for a subset of observations $R$ we have that $\forall r\in R\subset\left\{ 1,\dots,N\right\} :w_{r}=0$.
In this case, only the other observation count, those in the set $i\in\left\{ 1,\dots,N\right\} \backslash R$
for which the weights $w_{i}>0$ do not vanish. In this case, it is
immediately clear that the sample size going into the calculation
of $Var_{w}\left(y\right)$ is reduced, as the cases with zero weight
do contribute to the estimate of the mean. The sample size becomes
(at most) $N_{R}=N-\left\Vert R\right\Vert $, i.e., it is reduced
by the number of observations that are eliminated by the fact that
their weights are zero. The effective sample size is, assuming unit
weights for $i\notin R$ 
\[
N_{R}=\sum_{i=1}^{N}w_{i}=\left\Vert \left\{ 1,\dots,N\right\} \backslash R\right\Vert 
\]
which can be considered the effective sample size for a weighted sample
where some cases have unit weights ($w_{i}=1$), and the rest has
vanishing ($w_{r}=0$) weights.

In more general terms, the weights $w_{i}$ may be real valued and
positive, so that their relative variance 
\[
relvar\left(\boldsymbol{w}\right)=\frac{1}{K}\sum_{k=1}^{K}\left(\frac{w_{k}}{\overline{w}}-1\right)^{2}
\]
with $\overline{w}=\frac{1}{K}\sum_{k}w_{k}$ does not vanish. 

Consider a random variable $Y$ with finite mean $\mu$ and finite
variance $\sigma^{2}$. Let $y_{i}\in\mathbb{R},i=1,\dots,N$ denote
the observed values in a sample of size $K$ and assume the sample
requires weighting using the weights $w_{k}$. We are studying a weighted
estimator of the sample mean using weights $w_{k}\in\mathbb{R}^{+}$,
that is
\[
\overline{y}_{w}=M_{w}\left(y\right)=\frac{\sum_{k=1}^{K}w_{k}y_{k}}{\sum_{k=1}^{K}w_{k}}=\frac{1}{N}\sum_{k=1}^{K}\frac{w_{k}}{\overline{w}}y_{k}.
\]
Similarly, we can estimate the variance of $Y$
\[
Var_{w}\left(y\right)=\frac{\sum_{k=1}^{K}w_{k}\left(y_{k}-\overline{y}_{k}\right)^{2}}{N\overline{w}}=\frac{\sum_{k=1}^{K}w_{k}y_{k}^{2}}{N\overline{w}}-\left[\frac{\sum_{k=1}^{K}w_{k}y_{k}}{N\overline{w}}\right]^{2}=\overline{y_{w}^{2}}-\left(\overline{y}_{w}\right)^{2}.
\]

The weights $w_{k}$ can be chosen to reflect unequal sampling probability.
For example, if the probability of sampling observation $i$ is given
by $P_{i}>0,$the weights could be chosen as $w_{i}=P_{i}^{-1}$ to
counter unequal probabilities of being included in the sample. 

Another scenario is that the $y_{i}$ are statistics based on smaller
samples taken from certain well-defined groups (sampling units, sub-populations,
etc.). In that case, $y_{k}=f_{k}\left(y_{k1},\dots,y_{kN_{k}}\right)$
are (potentially also weighted) aggregates based on the within group
observations $y_{k1},\dots,y_{kN_{k}}$. In this case, each of the
groups may receive a different weight to account for and correct discrepancies
between expected population proportion vs. sample sizes for each group.

Two things are worth noting here:
\begin{enumerate}
\item The variance estimate $Var_{w}\left(y\right)$ under non-uniform weights
may not be the same as the variance calculated under simple random
sampling $Var_{1}\left(y\right)$ (where every observation is 'worth
the same', as all $w_{i}=1$) due to the weights being used. 
\item The quality of the variance estimator $Var_{w}\left(y\right)$, may
differ from the quality of the unweighted estimate $Var_{1}\left(y\right).$
The term quality is chosen to be deliberately vague here. One well
known issue with weighting is that it improve one quality (reduce
bias) at the cost of another (increase uncertainty). 
\end{enumerate}
The well cited general formula for the effective sample size is related
to this cost of weighting. The expression to calculate the effective
sample size was first provided by Kish (1965) and then given in simplified
form by Kish (1992). It can be derived by examining the variance of
the weighted mean$\overline{y}_{w}$:

\[
Var\left(\overline{y}_{w}\right)=Var\left(\frac{\sum_{k=1}^{K}w_{k}y_{k}}{\sum_{k=1}^{K}w_{k}}\right)=\frac{\sum_{k=1}^{K}w_{k}^{2}Var\left(y_{k}\right)}{\left(\sum_{k=1}^{K}w_{k}\right)^{2}},
\]
which follows when assuming that the $y_{k}$ are independent. When
we additionally assume that $\forall k:Var\left(y_{k}\right)=\sigma^{2}$,
we obtain
\[
Var\left(\overline{y}_{w}\right)=\frac{\sum_{k=1}^{K}w_{k}^{2}\sigma^{2}}{\left(\sum_{k=1}^{K}w_{k}\right)^{2}}=\sigma^{2}\frac{\sum_{k=1}^{K}w_{k}^{2}}{\left(\sum_{k=1}^{K}w_{k}\right)^{2}}=\frac{\sigma^{2}}{N}\frac{\overline{w^{2}}}{\overline{w}^{2}}.
\]

Using the same notation as above, Kish's (1965) estimate of effective
sample size is given by
\[
n_{eff}\approx\frac{\left(\sum_{k=1}^{K}w_{k}\right)^{2}}{\sum_{k=1}^{K}w_{k}^{2}}\rightarrow Var\left(\overline{y}_{w}\right)=\frac{\sigma^{2}}{n_{eff}}.
\]

Across the literature, there is another expression that is often given
for the effective sample size. This expression is based on the relvariance
of the weights. That is 
\[
D_{eff}=1+cv^{2}\left(\boldsymbol{w}\right)
\]
where 
\[
1+cv^{2}\left(\boldsymbol{w}\right)=1+Var\left(\frac{\boldsymbol{w}}{\overline{w}}\right)=1+\frac{1}{N}\sum_{i=1}^{N}\left(\frac{w_{i}}{\overline{w}}-1\right)^{2}
\]
since $M\left(\boldsymbol{w}\right)=\overline{w}$. Then we have
\[
1+cv^{2}\left(\boldsymbol{w}\right)=\frac{1}{N\overline{w}^{2}}\sum_{k=1}^{K}w_{k}^{2}=N\frac{\sum_{k=1}^{K}w_{k}^{2}}{\left(\sum_{k=1}^{K}w_{k}\right)^{2}}=\frac{N}{n_{eff}}=D_{eff}.
\]

\section{Satterthwaite's Formula implies Kish's $n_{eff}$}

For a sample $X_{1},...,X_{K}$ with weights $w_{1},\dots,w_{K}$
of $K$ observations, consider
\[
\nu^{*}\approx\frac{\left(\sum_{k=1}^{K}w_{k}S_{k}^{2}\right)^{2}}{\sum_{k=1}^{K}w_{k}^{2}\frac{S_{k}^{4}}{\nu_{k}}}
\]
which is the well known Satterthwaite estimator of the d.f. for the
variance estimate 
\[
S^{2}=\sum_{k=1}^{K}w_{k}S_{k}^{2}.
\]
The Kish (1965) approximation of the effective sample size can obtained
by assuming that all variance components are identical, i.e., $\forall i,j:S_{i}^{2}=S_{j}^{2}=S_{0}^{2}$.
In that case, we obtain
\[
\nu^{*}\approx\frac{\left(\sum_{k=1}^{K}w_{k}S_{0}^{2}\right)^{2}}{\sum_{k=1}^{K}w_{k}^{2}\left[S_{0}^{2}\right]^{2}}=\frac{\left[S_{0}^{2}\right]^{2}\left(\sum_{k=1}^{K}w_{k}\right)^{2}}{\left[S_{0}^{2}\right]^{2}\sum_{k=1}^{K}w_{k}^{2}}=n_{eff}.
\]

In this case, no small $K,\nu_{k}$ is needed as the assumption made
by Kish eliminates the need to insert estimates of $E\left(S^{4}\right).$The
result turns out to be a simplified Satterthwaite-type estimator of
degrees of freedom if all component variances are known to be the
same, and only the weights are unequal.

\subsection{Comparing Kish, Satterthwaite, and the Corrected Version}

The next table shows a comparison of Kish 's effective sample size
and Satterthwaite's degrees of freedom (divided by component average
$\overline{\nu}$) for random weights$w_{k}\sim N(1,0.3)$. The last
three columns show ratios that indicate how close the estimates are
to the number of components $K$, in the case of Kish, or how close
these are to $K\overline{\nu}$, the expected value in the unweighted
case for equal component $\nu_{k}$.

\selectlanguage{english}%
\begin{table}[htbp]
\centering \caption{Results for different weights (random $w_{k}\sim N(1,0.3)$). Columns:
mean Kish, mean Satterthwaite (uncorrected), mean corrected Satterthwaite,
total sample size $K\overline{\nu}$, and ratios relative to $K$
or $K\overline{\nu}$.}
{\footnotesize{}%
\begin{tabular}{|c|c|r|r|r|r|r|r|r|}
\hline 
{\footnotesize$K$} & {\footnotesize$\overline{\nu}$} & {\footnotesize M(Kish)} & {\footnotesize M(Satt)} & {\footnotesize M(Corr)} & {\footnotesize$K\overline{\nu}$} & {\footnotesize Kish/$K$} & {\footnotesize Satt/$K\overline{\nu}$} & {\footnotesize Corr/$K\overline{\nu}$}\tabularnewline
\hline 
{\footnotesize 16} & {\footnotesize 1} & {\footnotesize 14.74} & {\footnotesize 6.16} & {\footnotesize 16.49} & {\footnotesize 16} & {\footnotesize 0.92} & {\footnotesize 0.39} & {\footnotesize 1.03}\tabularnewline
{\footnotesize 16} & {\footnotesize 5} & {\footnotesize 14.74} & {\footnotesize 55.36} & {\footnotesize 75.50} & {\footnotesize 80} & {\footnotesize 0.92} & {\footnotesize 0.69} & {\footnotesize 0.94}\tabularnewline
{\footnotesize 16} & {\footnotesize 50} & {\footnotesize 14.75} & {\footnotesize 712.31} & {\footnotesize 738.81} & {\footnotesize 800} & {\footnotesize 0.92} & {\footnotesize 0.89} & {\footnotesize 0.92}\tabularnewline
{\footnotesize 16} & {\footnotesize 500} & {\footnotesize 14.75} & {\footnotesize 7351.23} & {\footnotesize 7378.63} & {\footnotesize 8000} & {\footnotesize 0.92} & {\footnotesize 0.92} & {\footnotesize 0.92}\tabularnewline
\hline 
{\footnotesize 32} & {\footnotesize 1} & {\footnotesize 29.43} & {\footnotesize 11.28} & {\footnotesize 31.83} & {\footnotesize 32} & {\footnotesize 0.92} & {\footnotesize 0.35} & {\footnotesize 0.99}\tabularnewline
{\footnotesize 32} & {\footnotesize 5} & {\footnotesize 29.43} & {\footnotesize 107.84} & {\footnotesize 148.98} & {\footnotesize 160} & {\footnotesize 0.92} & {\footnotesize 0.67} & {\footnotesize 0.93}\tabularnewline
{\footnotesize 32} & {\footnotesize 50} & {\footnotesize 29.42} & {\footnotesize 1417.19} & {\footnotesize 1471.87} & {\footnotesize 1600} & {\footnotesize 0.92} & {\footnotesize 0.89} & {\footnotesize 0.92}\tabularnewline
{\footnotesize 32} & {\footnotesize 500} & {\footnotesize 29.43} & {\footnotesize 14659.25} & {\footnotesize 14715.88} & {\footnotesize 16000} & {\footnotesize 0.92} & {\footnotesize 0.92} & {\footnotesize 0.92}\tabularnewline
\hline 
{\footnotesize 64} & {\footnotesize 1} & {\footnotesize 58.78} & {\footnotesize 21.27} & {\footnotesize 61.81} & {\footnotesize 64} & {\footnotesize 0.92} & {\footnotesize 0.33} & {\footnotesize 0.97}\tabularnewline
{\footnotesize 64} & {\footnotesize 5} & {\footnotesize 58.78} & {\footnotesize 212.92} & {\footnotesize 296.08} & {\footnotesize 320} & {\footnotesize 0.92} & {\footnotesize 0.67} & {\footnotesize 0.93}\tabularnewline
{\footnotesize 64} & {\footnotesize 50} & {\footnotesize 58.81} & {\footnotesize 2830.18} & {\footnotesize 2941.39} & {\footnotesize 3200} & {\footnotesize 0.92} & {\footnotesize 0.88} & {\footnotesize 0.92}\tabularnewline
{\footnotesize 64} & {\footnotesize 500} & {\footnotesize 58.79} & {\footnotesize 29280.13} & {\footnotesize 29395.25} & {\footnotesize 32000} & {\footnotesize 0.92} & {\footnotesize 0.92} & {\footnotesize 0.92}\tabularnewline
\hline 
\end{tabular}}
\end{table}

\begin{table}[htbp]
\centering \caption{Results for equal weights (all $w_{k}=1$). Columns: mean Kish, mean
Satterthwaite (uncorrected), mean corrected Satterthwaite, total sample
size $K\overline{\nu}$, and ratios relative to $K$ or $K\overline{\nu}$.}
{\footnotesize{}%
\begin{tabular}{|c|r|r|r|r|r|r|r|r|}
\hline 
{\footnotesize$K$} & {\footnotesize$\overline{\nu}$} & {\footnotesize M(Kish)} & {\footnotesize M(Satt)} & {\footnotesize M(Corr)} & {\footnotesize$K\overline{\nu}$} & {\footnotesize Kish/$K$} & {\footnotesize Satt/$K\overline{\nu}$} & {\footnotesize Corr/$K\overline{\nu}$}\tabularnewline
\hline 
{\footnotesize 16} & {\footnotesize 1} & {\footnotesize 16.00} & {\footnotesize 6.45} & {\footnotesize 17.36} & {\footnotesize 16} & {\footnotesize 1.00} & {\footnotesize 0.40} & {\footnotesize 1.08}\tabularnewline
{\footnotesize 16} & {\footnotesize 5} & {\footnotesize 16.00} & {\footnotesize 59.31} & {\footnotesize 81.03} & {\footnotesize 80} & {\footnotesize 1.00} & {\footnotesize 0.74} & {\footnotesize 1.01}\tabularnewline
{\footnotesize 16} & {\footnotesize 50} & {\footnotesize 16.00} & {\footnotesize 771.06} & {\footnotesize 799.90} & {\footnotesize 800} & {\footnotesize 1.00} & {\footnotesize 0.96} & {\footnotesize 1.00}\tabularnewline
{\footnotesize 16} & {\footnotesize 500} & {\footnotesize 16.00} & {\footnotesize 7970.16} & {\footnotesize 8000.04} & {\footnotesize 8000} & {\footnotesize 1.00} & {\footnotesize 1.00} & {\footnotesize 1.00}\tabularnewline
\hline 
{\footnotesize 32} & {\footnotesize 1} & {\footnotesize 32.00} & {\footnotesize 11.93} & {\footnotesize 33.78} & {\footnotesize 32} & {\footnotesize 1.00} & {\footnotesize 0.37} & {\footnotesize 1.06}\tabularnewline
{\footnotesize 32} & {\footnotesize 5} & {\footnotesize 32.00} & {\footnotesize 116.40} & {\footnotesize 160.96} & {\footnotesize 160} & {\footnotesize 1.00} & {\footnotesize 0.73} & {\footnotesize 1.01}\tabularnewline
{\footnotesize 32} & {\footnotesize 50} & {\footnotesize 32.00} & {\footnotesize 1540.29} & {\footnotesize 1599.91} & {\footnotesize 1600} & {\footnotesize 1.00} & {\footnotesize 0.96} & {\footnotesize 1.00}\tabularnewline
{\footnotesize 32} & {\footnotesize 500} & {\footnotesize 32.00} & {\footnotesize 15938.14} & {\footnotesize 15999.89} & {\footnotesize 16000} & {\footnotesize 1.00} & {\footnotesize 1.00} & {\footnotesize 1.00}\tabularnewline
\hline 
{\footnotesize 64} & {\footnotesize 1} & {\footnotesize 64.00} & {\footnotesize 22.75} & {\footnotesize 66.24} & {\footnotesize 64} & {\footnotesize 1.00} & {\footnotesize 0.36} & {\footnotesize 1.04}\tabularnewline
{\footnotesize 64} & {\footnotesize 5} & {\footnotesize 64.00} & {\footnotesize 230.86} & {\footnotesize 321.20} & {\footnotesize 320} & {\footnotesize 1.00} & {\footnotesize 0.72} & {\footnotesize 1.00}\tabularnewline
{\footnotesize 64} & {\footnotesize 50} & {\footnotesize 64.00} & {\footnotesize 3078.86} & {\footnotesize 3200.01} & {\footnotesize 3200} & {\footnotesize 1.00} & {\footnotesize 0.96} & {\footnotesize 1.00}\tabularnewline
{\footnotesize 64} & {\footnotesize 500} & {\footnotesize 64.00} & {\footnotesize 31874.41} & {\footnotesize 31999.90} & {\footnotesize 32000} & {\footnotesize 1.00} & {\footnotesize 1.00} & {\footnotesize 1.00}\tabularnewline
\hline 
\end{tabular}}
\end{table}

\selectlanguage{american}%
It can be seen that with increasing $\overline{\nu}\rightarrow\infty$
the Satterthwaite and the Kish estimates converge. This is expected
as Kish's estimator is obtained by ignoring the component-wise degrees
of freedom $\nu_{k}$. Therefore, Kish's approximation should only
be applied when it is known that all components estimate the same
variance, and in addition, it is known that the effect of estimating
the component variances can be ignored. 

\section{Examples Beyond Effective Sample Size}

\subsection{Jackknifing }

Let $T=T\left(X_{1},...,X_{L}\right)$ be a statistic that is computed
based on $l=1,...,L$ samples $X_{l}$. The jackknife estimate (e.g.
Efron \& Stein, 1981) of the variance $Var\left(T\right)$ is based
on calculating so-called pseudo-values 
\[
T_{k}=T\left(X_{1},\dots,X_{k-1},X_{k+1},\dots,X_{K}\right)
\]
which compute the same statistic under omission of the observation
$X_{k}$. Then, the variance of the statistic $T$ can be estimated
as 
\[
Var\left(T\left(X_{1},\dots,X_{K}\right)\right)=\frac{K-1}{K}\sum_{k=1}^{K}\left[T_{k}-T_{\cdot}\right]^{2}
\]
with $T_{\cdot}=\frac{1}{K}\sum_{k}T_{k}$

Each of the terms represents a variance component $S_{k}^{2}=\frac{K-1}{K}\left[T_{k}-T_{\cdot}\right]^{2}$and
$Var\left(T\right)=S^{2}=\sum S_{k}^{2}$ . The constant weight factor
$\frac{K-1}{K}$ is eliminated when forming the Satterthwaite expression,
and we can assume $\nu_{k}=1$ for all $k$. Hence, we obtain
\[
\nu^{*}\approx\frac{3\left(\sum_{k=1}^{K}\left[T_{k}-T_{\cdot}\right]^{2}\right)^{2}}{\sum_{k=1}^{K}\left[T_{k}-T_{\cdot}\right]^{4}}-2
\]

since $\nu_{k}+2=3$ for all $k.$ 

\subsection{Total Variance under Multiple Imputations}

The total variance of an estimator when imputations are used is a
prime example of a weighted variance estimate. The total variance
is defined as the combination of sampling and imputation variance,
and is given by
\[
Var\left(total\right)=Var\left(sampling\right)+\frac{M+1}{M}Var\left(imputation\right)
\]
according to Rubin (1987). Rubin \& Schenker (1986) discuss this case
and propose an approximation of the effective d.f. by assuming the
sampling variance has infinite d.f., while the imputation variance
has finite d.f. However, the sampling variance may be estimated according
to some resampling scheme (Johnson \& Rust, 1992) and the imputation
variance is the sample variance across $M$ imputation based calculations
of the same statistic. Lipsitz et al. (2002) provide an improved approximation
that also takes the finite d.f. of the sampling variance into account.
In this case, the following weights can be used

\[
\left(w_{1},w_{2}\right)=\left(1,\frac{M+1}{M}\right)
\]
and the corrected equation can be directly applied. That is 
\[
\nu_{total}^{*}\approx\frac{\left(Var\left(sampling\right)+\frac{M+1}{M}Var\left(imputation\right)\right)^{2}}{\frac{Var\left(sampling\right)^{2}}{\nu_{sampling}+2}+\left(\frac{M+1}{M}\right)^{2}\frac{Var\left(imputation\right)^{2}}{\nu_{imputation}+2}}-2
\]
Note that $\nu_{imputation}=M-1$ and that $\nu_{sampling}$ may require
estimation using, for example, the corrected Satterthwaite equation
applied to the jackknife variance components as illustrated in the
previous subsection.

\subsection{A Corrrected Welch Test $d.f.$}

The modified Welch (1947) test uses the d.f. $\nu_{1}=N_{1}-1$ and
$\nu_{2}=N_{2}-1$ of each of the variances of two samples to pool
the variance. The Welch-Satterthwaite formula used in this case can
be adjusted using the corrected Satterthwaite formula derived above.
Then we have 
\[
\nu^{*}\approx\frac{\left(\frac{1}{N_{1}}S_{1}^{2}+\frac{1}{N_{2}}S_{2}^{2}\right)^{2}}{\frac{S_{1}^{4}}{N_{1}^{2}\left(v_{1}+2\right)}+\frac{S_{1}^{4}}{N_{2}^{2}\left(v_{2}+2\right)}}-2
\]
where $S_{k}^{2}$ are the sample variances and weights 
\[
w_{k}=\frac{1}{N_{k}}
\]
in this case, for the d.f. $\nu_{*}$ of the sample size weighted
pooled variance used in the Welch test.

\section{Conclusion}

The approaches presented here were developed under different constraints.
The Kish (1965, 1992) effective sample size puts the strongest constraints
on the estimator, as it has been developed as tool for a simple approximation
when only the weights are known and all components of a variance estimate
can be assumed to have the same expectation, and each component is
a well estimated variance with moderate to large degrees of freedom. 

The Satterthwaite (1941, 1946) estimate of the effective degrees of
freedom was developed for weighted sums of variance components with
varying degrees of freedom. This is a much more general approach than
the simple approxiamtion provided by Kish (1965, 1992), but lacks
the ability to generate useful estimates when the component degrees
of freedom, $\nu_{k},$are small.

The corrected Satterthwaite effective degrees of freedom presented
here is based on von Davier's (2025a/b) prior research and ultimately
on the observation made by various authors that the original Welch-Satterthwaite
equation is heavily biased for small and moderate component degrees
of freedom (e.g. Boardman, 1974; Johnson \& Rust, 1992). In the current
paper, the prior improvements are superseded by a corrected version
of the equation that is solely based on the properties of expected
values of the fourth moment of a random variable. This corrected Welch-Satterthwaite
equation produces estimates that are much closer to the expected value,
when simulating data using the the ideal case, $K\overline{\nu}$,
also for small component $\nu_{k}$. It is calculated as
\[
\nu^{*}\approx\frac{\left[\sum_{k}w_{k}S_{k}^{2}\right]^{2}}{\sum_{k}w_{k}^{2}\frac{\left[S_{k}^{2}\right]^{2}}{\nu_{k}+2}}-2
\]
and with growing component degrees of freedom $\nu_{k}$ and components
$K$ it approaches the original Welch-Satterthwaite equation, while
for small $\nu_{k}$ it avoids the severe bias found with the original
formula.

\section*{References}

Boardman, T. J. (1974). Confidence Intervals for Variance Components
-{}- A Comparative Monte Carlo Study. Biometrics, 30(2), 251--262.
doi:10.2307/2529647

Efron, B. \& Stein, C. (1981). The jackknife estimate of variance.
The Annals of Statistics, pp. 586--596.

Johnson, E. G., \& Rust, K. F. (1992). Population Inferences and Variance
Estimation for NAEP Data. Journal of Educational Statistics, 17(2),
175--190. doi:10.2307/1165168

Kish, L. (1965). Survey sampling. John Wiley \& Sons. 

Kish, L. (1992). Weighting for unequal Pi. Journal of Official Statistics,
8(2), 183--200. 

Lipsitz, S., Parzen, M., \& Zhao, L. P. (2002). A Degrees-Of-Freedom
approximation in Multiple imputation. Journal of Statistical Computation
and Simulation, 72(4), 309--318. doi:10.1080/00949650212848

Rubin, D. B., \& Schenker, N. (1986). Multiple Imputation for Interval
Estimation From Simple Random Samples With Ignorable Nonresponse.
Journal of the American Statistical Association, 81(394), 366--374.
doi:10.2307/2289225

Satterthwaite, F. E. (1941). Synthesis of Variance. Psychometrika,
6(5), 309-316. doi:10.1007/BF02288586

Satterthwaite, F. E. (1946). An Approximate Distribution of Estimates
of Variance Components. Biometrics Bulletin, 2(6), 110--114. doi:10.2307/3002019

von Davier, M. (2025). An Improved Satterthwaite (1941, 1946) Effective
df Approximation. Journal of Educational and Behavioral Statistics,
0(0). doi:10.3102/10769986241309329

von Davier, M. (2025). An Improved Satterthwaite Effective degrees
of freedom correction for weighted syntheses of variance. arXiv. doi:10.48550/arXiv.2503.22080

Welch, B. L. (1947). The generalization of `Student’s’ problem when
several different population variances are involved. Biometrika, 34(1/2),
28-35. doi:10.2307/2332510
\end{document}